\documentclass{appolb}%
\usepackage{graphicx}
\usepackage{amsmath}

\def\vec#1{\ensuremath{\mathchoice
                     {\mbox{\boldmath$\displaystyle#1$}}
                     {\mbox{\boldmath$\textstyle#1$}}
                     {\mbox{\boldmath$\scriptstyle#1$}}
                     {\mbox{\boldmath$\scriptscriptstyle#1$}}}}

\begin{document}
%
\title{A relativistic model for the Charmonium spectrum with a reduced number of free 
parameters
}
\author{M. De Sanctis \footnote{mdesanctis@unal.edu.co}
\address{Universidad Nacional de Colombia, Bogot\'a, Colombia }
\\
}
\maketitle
\begin{abstract}
A previously introduced reduction of the Dirac equation is used
to study the Charmonium spectrum.
A regularized vector potential that only depends on the coupling constant and on the regularization radius is adopted, considering the interacting quark as an extended source of the Chromo-electric field. 
A scalar interaction is also introduced with some constraints for its parameters.
A good description of the  structure of the Charmonium spectrum is obtained with only three free parameters.
\end{abstract}
\section{Introduction}
In a previous work  \cite{localred}  that will be denoted here as I,  we studied a relativistic reduction of the Dirac equation for quark composed systems. 
In that work we analyzed the theoretical fundaments of  that reduced equation
and showed that it  was able to reproduce the Charmonium
spectrum with high accuracy but using a relatively large number of free parameters.
More precisely, we used eight or nine parameters to reproduce the spectrum,
taking into account a possible energy-dependence of the interaction.

On the contrary, in the present work, by using the same reduced 
relativistic equation we shall try 
to reproduce the main structure of the  Charmonium spectrum with a very small number of parameters, possibly with evident physical meaning.
To this aim we shall \textit{determine} two parameters of the model in order to reduce
the total number of free parameters.

As in I, we use a vector-scalar interaction model to represent the dynamics of Charmonium. 

\noindent
In more detail, we shall use a specific form of the vector interaction
(possibly related to QCD)
 that corresponds to a regularized Color interaction of the quarks.
 To this aim, we assume a non-pointlike
distribution of their Chromo-electric charge. 
This model has  been  studied in detail in the work \cite {chromomds} that will be denoted as II in the following. \\
To avoid repetition, the reader will be directed  to the specific parts of
the works I and II, when necessary.
In those works, the reader can also find the references on which 
the whole study is based.
Recalling that the aim of this work is to study  the Charmonium spectroscopy
with a \textit{relativistic} model with only \textit{three free parameters},
we briefly mention  below (with no attempt of completeness), some relatively recent studies on this subject.\\
We first quote the nonrelativistic models.
In Ref. 
\cite{Wei-Jun} 
two  models with five parameters are constructed to study the Charmonium resonances and their electromagnetic transitions.
A momentum-helicity  model 
\cite{Radin}, 
also with five parameters, was
proposed for the Charmonium spectrum.
A model for the spectrum and decay rates with a Coulomb-like potential,
a linear confining potential and a potential derived from the instanton vacuum
\cite{Praveen}, 
was studied. This model used four parameters.
The instanton effects are also studied in another model with six parameters
\cite{Ulugbek}.
The mass spectrum was calculated in the framework of nonrelativistic QCD, with seven parameters
\cite{Raghav}.
The Charmonium properties were studied by solving the Schr\"odinger equation 
with the discrete variable representation method
\cite{Bhaghyesh}. This last model used five parameters.\\
A semirelativistic model with a Coulomb plus linear potential using five parameters \cite{Viren} was studied.\\
As for the fully relativistic models we recall the studies performed by means
the Covariant Spectator Theory, with  vector, scalar and pseudoscalar interactions
\cite{Sofia1,Alfred}.
The authors  studied, by means of the same model, heavy and heavy-light mesons.
They used three free parameters
and a fixed cutoff parameter to regularize the momentum space integrals.
Other parameters are the constituent quark masses,
and the weight of
scalar and pseudoscalar coupling for the confining interaction.
The authors analyzed the dependence of the results on these last parameters by means of 
 different calculations in which they are
considered as fixed or as free parameters.\\
Another relativistic model was based on the use of a momentum space integral
equation with positive energy Dirac spinors.
A complete one-gluon exchange interaction with other phenomenological scalar terms
were used
\cite{mdsfch}.
In that work two different potentials were considered, with seven and eight free parameters, respectively.
The same model was applied to the study of the Bottomonium spectrum
\cite{mdsfbo}.\\
A covariant four-dimensional approach, based on the Schwinger-Dyson equations
with a vector contact interaction was used to study the first radial excitations of
heavy quarkonium
\cite{Bedo_Elena}.
For the Charmonium case, the authors used five parameters and fixed the
$c$-quark current mass at $1.09~ GeV$.
Furthermore, they obtained other results adjusting three parameters to obtain
the experimental mass of the $\eta'_c(2S)$.
The model was also generalized to study  the masses of light and heavy mesons and 
baryons
\cite{Guti_Elena}.\\
We conclude observing that the construction of a consistent model for the study
of Charmonium (and, more in general, of quarkonium) is still a very active field
of investigation.  

\vskip 0.5 truecm
\noindent
Going back to the present paper, its remaining content is organized as follows.\\
In Subsection \ref{symbnot} we briefly explain the symbols and the notation 
used in the work.\\
In Section \ref{diracred} we recall the main aspects of the reduced relativistic equation.\\
In Section \ref{genint} the general form of the interaction is introduced.\\
In Subsections \ref{vectint} and \ref{scalint} the details of the vector and scalar 
interactions are discussed, respectively.\\
Finally, the results of the model are presented and analyzed in Section \ref{charmres}.\\
Some conclusions are drawn in Section \ref{conclusions}.


\subsection{Symbols and Notation}\label{symbnot}
For the Dirac matrices (in the standard representation), 
for all the other operators and wave functions, 
we use the  notation introduced in I and also $\hbar=c=1$. 
%
We shall use
the generic word \textit{quark}  to denote both the $c$-quark and the antiquark $\bar c$
of the Charmonium system. The word antiquark will be used only when strictly 
necessary.
For the argument of the Color charge distribution, in Eqs. (\ref{vint}) and
(\ref{rhogauss}), we use $x=|\vec x|$.

\vskip 1.0 truecm

\section{The reduced Dirac equation}\label{diracred}
Following I, we summarize here the main aspects of the reduction of the Dirac equation
that is used in this work. 
The starting point is represented by the one-body reduction operator that, 
for the $i$-th
particle, takes the form:

\begin{equation}\label{defk}
K_i= K(m_i,E_i; \vec p_i, \vec \sigma_i)=
\begin{pmatrix} 1 \\ 
                {\frac {\vec \sigma_i \cdot \vec p_i} {m_i +E_i}  }
\end{pmatrix}~.
\end{equation}
where $m_i$, $E_i$, $\vec p_i$, $\vec \sigma_i$ respectively represent the mass, 
energy, momentum and Pauli matrix of the $i$-th constituent. 
The operator $K_i$, introduced in Eq. (21) of I, is applied to a two-component spinor and gives rise (for one particle) to a four-component vinculated Dirac spinor.
Note that $K_i$ is a \textit{local} operator, so that the complete equation is also local and can be solved in the coordinate space.\\
The two-body  reduced equation, given in Eq. (40) of I,  can be formally written as:
\begin{equation}\label{dir2red1}
 K_1^\dag \cdot K_2^\dag
(D_1 + D_2  +W_{(2)})
       K_1
 \cdot K_2
| \Phi> =0~.
\end{equation}
where $D_i$ ($i=1,2$) represents the standard operator of the free Dirac equation:
\begin{equation}\label{didef}
D_i=D(m_i,E_i; \vec p_i, \vec \alpha_i,\beta_i)=
\vec \alpha_i \cdot \vec p_i +\beta_i m_i -E_i~.  
\end{equation}
In Eq. (\ref{dir2red1}) we have  also introduced the two-body Dirac interaction operator 
$W_{(2)}$.\\
We consider the case of two equal mass particles $m_1=m_2=m$, in the 
Center of Mass (CM) reference frame, where the total momentum
$\vec P$ is vanishing.
In this frame the following relation holds:
\begin{equation}\label{momentumops}
\vec p_1= -\vec p, ~~ \vec p_2= \vec p
\end{equation}
where $\vec p$ represents the relative momentum operator, 
canonically conjugated
to the relative distance vector 
\begin{equation}\label{rdv}
\vec r =\vec r_2 - \vec r_1~.
\end{equation}
Furthermore, we assume that, in the CM, the two particles have the same energy:
\begin{equation}\label{eqen}
E_1=E_2= {\frac {E_T} {2}}= {\frac {M} {2}}  ~,
\end{equation}
where $E_T=M$ represents the mass off the resonant state.\\
In this way, 
we obtain the reduced equation in the form given by Eqs. (43) and (44) of I,
that is:

\begin{equation}\label{eq2redcm}
 \left[\left( 1 + {\frac {\vec p^2} {(E_T/2+m)^2} } \right)
\left( {\frac { 2 \vec p^2} {E_T/2+m} } +2m -E_T \right)
+ \hat W_{(2)} \right] | \Phi>=0
\end{equation}
where we have also introduced the 
two-body reduced interaction operator:

\begin{equation}\label{w2red}
\hat W_{(2)}= K_1^\dag \cdot K_2^\dag ~ W_{(2)}
~ K_1 \cdot K_2~.
\end{equation}
The expressions for the reduced scalar and vector two-body interactions
have been given in Eqs. (C.3) and (C.5) of I. \\
Eq. (\ref{eq2redcm}) is a local, energy-dependent equation,
free from continuum dissolution desease \cite{localred,such}, 
that can be advantageously used 
to study the spectroscopy of Charmonium and of other mesons.\\
We also recall that Eq. (\ref{eq2redcm}), being an energy-dependent effective equation, 
can be solved by means of the technique explained in detail in Section 7 of I.
%
\section{The general structure of the interaction}\label{genint}
As in Section 6 of I, 
for the two-body interaction $W_{(2)}$ that appears in
 Eqs. (\ref{eq2redcm}) and (\ref{w2red}),
we consider a standard sum of a vector and a scalar contribution, in the form:

\begin{equation}\label{wmodel}
W_{(2)}= W_{(2)}^v+ W_{(2)}^s~.
\end{equation}

\vskip 0.5 truecm
\noindent
For the vector interaction we take the following standard expression:
\begin{equation}\label{vvect}
W_{(2)}^v=V_{(2)}^v(r)
\gamma_1^0 \gamma_2^0 \cdot
\gamma_1^\mu \gamma_2^\nu g_{\mu \nu}
\end{equation}
%
%
\noindent
The potential function $V_{(2)}^v(r)$ will be discussed in the following
Subsection \ref{vectint}.\\
In order to have a local interaction operator, as explained in I,
 we have not included  retardation 
contributions,
consistently with
Eq. (\ref{eqen}):
we make the hypothesis that the quark energies $E_i=E_T/2$ are 
 \textit{fixed}; 
in other words, we assume that the quarks do not interchange energy 
with the effective gluonic vector field that mediates the interaction.  

\noindent
For the scalar interaction we take the expression:
\begin{equation}\label{vscal}
W^s_{(2)}=V_{(2)}^s(r)
\gamma_1^0 \gamma_2^0 ~.
\end{equation}
\textit{Many trials} have been performed to determine the specific form of
the potential functions $V_{(2)}^v(r)$ and $V_{(2)}^s(r)$
in order to reproduce the Charmonium spectrum with a very small number of free parameters.
In the two following subsections the specific properties of the two
potential functions will be discussed in detail.

\subsection{The vector interaction}\label{vectint}
The vector interaction is constructed according to the model proposed in 
the work II. 
In that model the quarks are considered as extended  sources 
of the Chromo-electric field.
In consequence, the Color charge distribution of these sources determines the
form of the potential and the value of the self-energy 
(\textit{i.e.}, the zero-point potential energy)
  that is \textit{not}
introduced as an extra parameter. \\
Considering the objective of the present work,
the contents of the Section 2 of II 
can be syntetically rewritten  
in the following way.\\
Given a Color charge distribution $\rho(x)$ (obviously rotationally invariant)
for each quark,
the  attractive interaction energy between the quark and antiquark has the form:
\begin{equation}\label{vint}
 V^{int}( r)= 
- {\frac 4 3} \alpha_v
\int d^3x \int d^3 x' 
\rho( x) \rho( x') 
{\frac {1} {| \vec x - \vec x' + \vec r|}}~.
\end{equation}
(This interaction energy was denoted as $W^{int}_{q \bar q}$ in II.)  
Now we recall that $4/3$ is the Color factor for the quark-antiquark interaction, 
$\alpha_v$ is the Color (vector) effective coupling constant (more frequently
denoted as $ \alpha_{strong}$).
Note that the last factor of Eq. (\ref{vint}) represents the Coulombic term.
Due to the presence of that term, the interaction energy $V^{int}( r)$ is also
Coulombic at large distance.\\
As shown in II, the Color charge distributions give rise to a positive 
zero-point self-energy that will be denoted as $\bar V_v$ in the present work
(while the same quantity was defined  $W^{self}$ in II).
It is given by  the following relation:
\begin{equation}\label{vbar}
\bar V_v= -V^{int}(r=0)~.
\end{equation}
Furthermore, we note that the time component of the vector interaction,
studied in II, corresponds to $V_{(2)}^v(r)$ of Eq. (\ref{vvect}).
In conclusion, for this quantity, we have:
\begin{equation}\label{vv2}
V^v_{(2)}( r)= \bar V_v + V^{int}( r)~.
\end{equation}
The Color charge distributions of the quarks \textit{regularize}
the interaction potential at $ r =0$ and produce the self-energy $\bar V_v$.
As a result, we obtain for $V^v_{(2)}( r) $ a potential that is \textit{vanishing} at
 $r =0$ and approaches the maximum value $\bar V_v$ (with a Coulombic behavior)
as $r\rightarrow \infty$.\\

\vskip 0.5 truecm
\noindent
The best reproduction of the experimental data has been  obtained with a Gaussian
Color charge distribution, of the form:
\begin{equation}\label{rhogauss}
\rho( x)={\frac {1} {(2 \pi d^2)^{3/2} }}
\exp\left(-{\frac {\vec x^2} {2d^2} } \right)~.
\end{equation}
With this distribution, $\bar V_v$   and $V^{int}( r)$ can be
calculated analitically. The results are:
\begin{equation}\label{vbargauss}
\bar V_v={\frac 4 3} {\frac {\alpha_v}  {d}}{\frac {1}  {\sqrt{\pi}}   }
\end{equation}
and
\begin{equation}\label{vintgauss}
 V^{int}( r)=- {\frac 4 3}  {\frac {\alpha_v} {r}}
\text{erf} \left(  {\frac {r} {2d}  }\right)~.
\end{equation}
The same regularization function  
$F_v(r)=\text{erf} \left(  {\frac {r} {2d}  }\right)$ 
was also used in Eq. (61) of I, with $2d=d_v$.
The relevant properties of this regularization function are also explained there.\\
Finally, note that $\bar V_v$  is \textit{not} a free parameter 
but is determined by $\alpha_v$ and $d$ that represent the only
free parameters of the vector interaction.
 
\subsection{The scalar interaction and an additional constraint}\label{scalint}
In order to reproduce with a reasonable accuracy the experimental data of the Charmonium spectrum, we have verified that it is strictly necessary to introduce  a scalar interaction.\\
However, in the present context, it has not been possible to construct 
a more fundamental model to represent this interaction.

\vskip 0.5 truecm
\noindent
 After trying different forms for $V_s(r)$, we have found that a \textit{negative}
function, regular at $r=0$, that goes to zero as $r\rightarrow \infty$,
is needed to reproduce the spectrum.
The simplest expression, with only two free parameters, is a Gaussian function:
\begin{equation}\label{vsgauss}
V_s^G(r)= -\bar V_s \exp 
\left(- {\frac { r^2} {r_s^2} } \right)~.
\end{equation}
We also report, in the results of the following section, 
a \textit{very simple test} with a constant
interaction:
\begin{equation}\label{vsconst}
V_s^C(r)= -\bar V_s~.
\end{equation}
In this case, the reproduction of the spectrum is obviously \textit{worse}
than that obtained with $V_s^G(r)$.\\
Furthermore we have used a \textit{two region}  potential,
studying the possibility that the scalar interaction is related to
the interchange of a scalar particle of mass $m_b$.\\
After trying different parametrizations, we found that, in any case, it is necessary
to consider two spatial  regions: an inner region, with a relatively soft 
depenence on $r$ and an outer region in which the scalar interaction is represented by
a standard Yukawa function related to the interchange of a mass $m_b$.
The expression that has been used has the following form:
 
\begin{equation}\label{vstworeg}
 V_s^T(r)= \begin{cases}
  - \bar V_s\left[1-b({\frac {r} {r_s}})^{^p} \right] ~\text{for} ~r\leq r_s,\\
  - {\frac {\beta} {r}} \exp(- {\frac {r} {r_b}}) ~~\text{for} ~r> r_s ~.
\end{cases}
\end{equation}
In the previous expression we use the same symbol $r_s$ introduced for
the Gaussian potential.
But now it has a different meaning:
it fixes the limit between the inner and outer spatial regions;
$b$ and $p$ are adimensional parameters 
whose numerical value will be given in the next section, $\beta$ represents the
adimensional coupling constant of the Yukawa interaction and, finally, 
$r_b=1/m_b $.\\
We require that $V_s^T(r)$ and its first derivative are \textit{continous} at $r=r_s$.
These two conditions respectively give:
\begin{equation}\label{betafix}
\beta= {\bar V_s }r_s (1-b)\exp( {\frac {r_s} {r_b}})
\end{equation}
and
\begin{equation}\label{r_bfix}
r_b=r_s{\frac {1-b} {bp-1+b} }~.
\end{equation}
Further discussions about the scalar interaction and its specific form
$V_s^T(r)$ are postponed to the next section.
\vskip 0.5 truecm
\noindent
Finally, in order to reduce the number of free parameters of the scalar interaction,
we  introduce  a phenomenological constraint on the parameter $\bar V_s$.
Recalling the discussion of the previous subsection about the self-energy of the vector interaction, analogously to Eq. (\ref{vbar}),
we assume here that, 
for the scalar potentials of Eqs. (\ref{vsgauss}), (\ref{vsconst})
and (\ref{vstworeg}), 
$~V_s(r=0)=-\bar V_s$
represents the (negative) self-energy of the scalar interaction.\\
Considering this starting point,
we shall use the following phenomenological \textit{balance} equation:
\begin{equation}\label{balance}
\bar V_v =  2 m_q - \bar V_s~.
\end{equation}
It means that the self-energy of the vector interaction 
equals the rest energy of the quarks plus the negative self-energy of the
scalar interaction. \\
In this way, solving  Eq. (\ref{balance}) with respect to $\bar V_s$,  we avoid to introduce this quantity as a free parameter;
on the contrary, it is \textit{determined} by the other parameters of the model.\\
Some more comments wil be given in the following Section \ref{charmres}
when analyzing the results of the calculation.


\section{The result for the Charmonium spectrum }\label{charmres}
In this section we apply the model to study the Charmonium spectrum
with the interaction introduced in the previous  Section \ref{genint}.\\
The relativistic, energy-dependent  equation (\ref{eq2redcm})    is  
solved with the same  technique explained in Section 7 of I, to which we refer the reader.\\
We use a variational basis of harmonic oscillator (HO) wave functions 
that, in the coordinate space, have the form given in Eq. (63) of I :
\begin{equation}\label{basisho}
\Phi_{n; L,S,J}(\vec r)=< \vec r|  n; L, S, J> =
R_{n,L}(r;\bar r) {[Y_L(\hat r) \otimes \chi_S]}_J ~.
\end{equation}
In the previous equation the trial radial function is represented by
$R_{n,L}(r;\bar r)$, being $n$ the principal  HO quantum number and
$\bar r$ the variational parameter with the dimension of longitude;
 $Y_{L,M_L}(\hat r)$ is the corresponding spherical harmonic
and $\chi_{S,M_S}$, with $S=0,1$ is the  $c~ \bar c$
coupled spin function.
The orbital angular momentum  and the spin are standardly coupled to 
the total angular momentum $J,M_J$.
For brevity we do not write $M_J$ because it is unrelevant for the calculations
of rotationally scalar operators.\\
Finally,
for simplicity reasons, we do not consider the possibility of mixing between 
states with different values of $L$, because these effects are usually considered negligible in these calculations.\\
The analytic form of the radial HO functions is given in Eq. (64) of I.

\vskip 0.5 truecm
\noindent
As for the fit procedure,
we have determined the free parameters of the model by minimizing the quantity
\begin{equation}\label{leastsquare}
D^2=\sum_i(E_i^{th.} -M_i^{exp} )^2
\end{equation}
where $E_i^{th.}$ and $M_i^{exp} $ respectively represent the calculated
energy and the experimental mass (rest energy) of the i-th resonance.

\noindent
Due to the higher number of parameters (more precisely, eight or nine) used in I,
in that work a better reproduction of the spectrum was obtained.
On the other hand,
we try here to fit the whole spectrum with \textit{three} parameters only.\\
To this aim we also fix the mass of the  $c$-quark  at the value given 
by the PDG as the
  ``running" mass in the $\overline{MS}$ scheme.
This value is presently $m_q =1.27~ GeV$ \cite{pdg}.

\vskip 0.5 truecm
\noindent
All the results for the spectrum are given in Table \ref{tabres}.\\
In last column of this table we give the experimental values of the resonances.
In particular,
we have considered all the eight experimentally observed resonances,
whose energies are below the open charm threshold  $D~\bar D$;
 we have also taken eight
not controversial resonances at higher energies.
For a discussion about the phenomenological interpretation of the resonances
in different models, the interested reader is referred to Ref. \cite{mdsfch}. \\
In the columns denoted by ``Gauss", ``Const." and ``Two Reg." we report the
theoretical results given respectively by the Gaussian scalar potential
of Eq. (\ref{vsgauss}), by the constant scalar potential of Eq. (\ref{vsconst})
and by the \textit{two region} scalar potential of Eq. (\ref{vstworeg}).\\
Considering Eq. (\ref{leastsquare}),
we report, for simplicity, in the last line of Table \ref{tabres} the quantity
$Q=D^2/100$ en $MeV^2$ in order to give an indication about the quality 
of the fit for the three scalar potentials.

\vskip 0.5 truecm
\noindent
The values of the parameters of the model are shown in Table \ref{tabpar}.\\
In particular,
we give the values obtained with the fit procedure for the independent parameters:
 the effective coupling constant $\alpha_v$ 
and the regularization radius $d$ for the effective vector interaction, and
the radius $r_s$    of the scalar interaction. 
We also give the values of dependent parameters $\bar V_v$
of Eq. (\ref{vbargauss})
 and $\bar V_s $,
determined by means of Eq. (\ref{balance}).\\
The same notation (for the different scalar potentials) as in Table \ref{tabres}
is used in Table \ref{tabpar}.

\vskip 0.5 truecm
\noindent
The main results of this work are those obtained with the Gaussian potential $V_s^G(r)$ of Eq. (\ref{vsgauss}) for the scalar interaction. 
As shown in Table \ref{tabres} a good overall reproduction of the spectrum 
is obtained with only \textit{three} free parameters.

\vskip 0.5 truecm
\noindent
As anticipated in Subsection \ref{scalint}, we have also tried to reproduce the spectrum (as a very simple test) with  a constant scalar potential. 
In this case we need only two free parameters: $\alpha_v$ and $d$. \\
The corresponding value of $Q$ in Table \ref{tabres} shows that the quality of the fit is considerably worse than that given
by the Gaussian scalar potential.

\vskip 0.5 truecm
\noindent
Finally, with the \textit{two region} potential,
we have  explored the possibility that the scalar interaction is given, at least in the outer region,  by the standard exchange of a scalar particle,
by using the potential $V_s^T(r) $  of Eq. (\ref{vstworeg}), 
with the continuity conditions
of Eqs. (\ref{betafix}) and (\ref{r_bfix}).\\
For this case, we point out that,
after some trials,
we have fixed (for simplicity) the power $p$ of the inner part of the potential 
at $p=3/2$
and the parameter $b$ at the value $b=2^{-1/2}$. 
We have verified that no significant improvement is obtained 
varying these  values.

\noindent
We also give the values of other \textit{dependent} parameters of this interaction:

\noindent
i) for the range of the Yukawa interaction  $r_b$,
fixed by Eq. (\ref{r_bfix}),
 we have obtained $r_b=0.7594~fm$,
corresponding to a scalar boson mass of $m_b =0.2598~GeV$;\\
ii) for the Yukawa coupling costant $\beta$, fixed by Eq. (\ref{betafix}),
 we  have found
$\beta=29.75~$.\\
Considering the results of Table \ref{tabres},
we observe that no significative improvement for the Charmonium spectrum is obtained with respect
to the case of the Gaussian scalar potential.
Furthermore, the radius of the inner region $r_s$ is large with respect to
the range $r_b$ of the (hypothetical) Yukawa interaction.\\
In conclusion,
the model does not show clear evidence for the exchange of a scalar particle.

\vskip 0.5 truecm
\noindent
Finally, for
all the scalar interactions we have also tried to consider $\bar V_v$ and $\bar V_s$
as free parameters, ignoring Eqs.  (\ref{vbargauss}) and (\ref{balance}),
but no significant improvement in the reproduction of the spectrum has been obtained.\\
As in I, we have also used the   reduced equation obtained by the relativistic
Mandelzweig and Wallace equation  \cite{mwa,mwb} ;
the obtained results are \textit{very similar} to those discussed above (obtained 
by using the reduced Dirac-like Equation  (\ref{eq2redcm})).
For this reason, they have not been shown in the  paper.\\
Some more comments are given in the Conclusions.
\section{Conclusions}\label{conclusions}
In this paper we have shown that a relativistic, energy-dependent, local equation
can be used to describe the Charmonium spectrum with good accuracy using only three
free parameters.
A standard mixture of a vector and scalar interaction has been considered.
As for the vector part of the interaction, the regularization radius fixes 
the quark self-energy that is determined in this way as a \textit{dependent} parameter.\\
For the scalar interaction  a phenomenological Gaussian potential is taken;
the possibility of  the interchange of a scalar particle has been also explored.
A balance equation is used to determine  the value of the scalar potential at $r=0$.\\
Furher investigation is needed to understand in more detail the nature and the origin of the scalar interaction.

\vskip 0.5 truecm
\centerline{{\bf Acknowledgements}}
The author thanks the group of  ``Gesti\'on de Recursos de Computo Cient\'ifico,
Laboratorio de Biolog\'ia Computacional,
Facultad de Ciencias - Universidad Nacional de Colombia"
for the access to the computation facilities that were used to perform the numerical calculations 
of this work.




\newpage
\begin{table*}

\caption{Comparison between the experimental average values \cite{pdg} 
 of the Charmonium spectrum (last column)
and the theoretical results of the model.
All the masses are in MeV. 
The quantum numbers  $n$, $L$, $S$ and $J$ have been introduced 
in Eq. (\ref{basisho});
they represent the principal quantum number, the orbital angular momentum, the spin 
and the total  angular momentum, respectively.
The results of the columns Gauss, Const. and Two Reg. refer to the different forms
of the scalar interaction, 
as specified in the text. 
A line divides the resonances below and above the open Charm threshold.
At the bottom, the quantity $Q$ gives an indication of the quality of the fit, as explained in the text.  }

\begin{center}
\begin{tabular}{cccccc}
\hline
\hline \\
Name & $n^{2S+1}L_J$ & Gauss & Const.  &  Two Reg. & Experiment          \\
\hline \\
$\eta_c$    &  $1^1 S_0 $     & 2989 & 3007   & 2990    & 2983.9   $\pm$  0.5   \\
$J/\psi$    &  $1^3 S_1 $     & 3092 & 3100   & 3092    & 3096.9   $\pm$  0.006 \\
$\chi_{c0}$ &  $1^3 P_0 $     & 3420 & 3386   & 3419    & 3414.71  $\pm$ 0.30   \\
$\chi_{c1}$ &  $1^3 P_1 $     & 3499 & 3461   & 3497    & 3510.67  $\pm$ 0.05   \\
$ h_c$      &  $1^1 P_1 $     & 3511 & 3464   & 3509    & 3525.38  $\pm$ 0.11   \\ 
$\chi_{c2}$ &  $1^3 P_2 $     & 3565 & 3556   & 3562    & 3556.17  $\pm$ 0.07   \\
$\eta'_c$   &  $2^1 S_0 $     & 3649 & 3680   & 3643    & 3637.5   $\pm$ 1.1    \\
$\psi'$     &  $2^3 S_1 $     & 3680 & 3708   & 3673    & 3686.097 $\pm$ 0.010  \\
\\
\hline \\
$\psi(3770)$&     $1^3 D_1 $     & 3797 &  3756 & 3791  & 3778.1   $\pm$ 1.2  \\  
$\psi(3823)$&     $1^3 D_2 $     & 3832 &  3828 & 3826  & 3822.2   $\pm$ 1.2   \\
$\chi_{c1}(3872)$&$2^3 P_1 $     & 3893 &  3914 & 3890  & 3871.69  $\pm$ 0.17  \\
$\chi_{c2}(3930)$&$2^3 P_2 $     & 3928 &  3949 & 3926  & 3927.2   $\pm$ 2.6   \\
$\psi(4040)$&     $3^3 S_1 $     & 4014 &  4027 & 4020  & 4039     $\pm$ 1     \\
%
$\chi{c1}(4140)$& $3^3 P_1 $     & 4145 &  4135 & 4158  & 4146.8  $\pm$ 2.4    \\
$\psi(4230)    $& $4^3 S_1 $     & 4214 &  4195 & 4220  & 4218.7    $\pm$ 2.8      \\
$\chi{c1}(4274)$& $4^3 P_1 $     & 4272 &  4258 & 4269  & 4274    $\pm$ 7      \\
\\
\hline
\hline \\
$ Q       $&       &$ 22.6$& $138 $&$ 17.8  $& \\  
\\
\hline\\
~\\
~\\
~\\
~\\

\end{tabular}
\end{center}
\label{tabres}
\end{table*}



\begin{table*} 
\caption{Numerical values of the free and dependent parameters of the model; 
$m_q $ is fixed at the value of Ref. \cite{pdg}, as explained in the text. 
The reported numerical values represent the results of the fits of the free parameters
$\alpha_v$, $d$ and $r_s$; $\bar V_v$ and $\bar V_s$ are dependent parameters, as explained in the text. 
 }
\begin{center}
\begin{tabular}{lllll}
\hline 
\hline \\   
                &   &  & & Units     \\ 
\hline \\
%
 $m_q $   &  $1.27$     &            & $ $       &  GeV   \\
\hline \\   
                 &   Gauss   & Const.   &  Two Reg.  &           \\ 
\hline \\   
 $\alpha_v $&  $1.864  $ &$3.991  $ & $1.865  $    &         \\
 $d        $&  $0.1526 $ &$0.2665 $ & $0.1531 $    & fm      \\
 $r_s      $&  $1.879  $ &          & $1.991  $    & fm      \\     

\hline\\
$\bar V_v $&  $1.813 $ &$2.223 $ & $1.808 $    & GeV \\
$\bar V_s $&  $0.7268 $ &$0.3170 $ & $0.7321 $    & GeV \\
\hline

\end{tabular}
\end{center}

\label{tabpar}
\end{table*}

\end{document}